\documentclass{article}

\usepackage{arxiv}

\usepackage[utf8]{inputenc} % allow utf-8 input
\usepackage[T1]{fontenc}    % use 8-bit T1 fonts
\usepackage{hyperref}       % hyperlinks
\usepackage{url}            % simple URL typesetting
\usepackage{booktabs}       % professional-quality tables
\usepackage{amsfonts}       % blackboard math symbols
\usepackage{nicefrac}       % compact symbols for 1/2, etc.
\usepackage{microtype}      % microtypography
\usepackage{lipsum}
\usepackage{fancyhdr}       % header
\usepackage{graphicx}       % graphics
\graphicspath{{media/}}     % organize your images and other figures under media/ folder

\title{FL-CLEANER: byzantine and backdoor defense by CLustering Errors of Activation maps in Non-iid fedErated leaRning}

\author{
  Mehdi Ben Ghali \\
  Inserm, IMT Atlantique \\
  Brest, France\\
  \texttt{ben-ghali.ben-ghali@imt-atlantique.fr} \\
   \And
  Gouenou Coatrieux, Reda Bellafqira \\
  IMT Atlantique \\
  Brest, France\\
  \texttt{\{Gouenou.coatrieux, Reda.bellafqira\}@imt-atlantique.fr} \\
}

\usepackage{amsmath,amsfonts}
\usepackage{algorithmic}
\usepackage{algorithm}
\usepackage{multirow}
\usepackage{enumitem}

\begin{document}
\maketitle
\begin{abstract}
Federated Learning (FL) enables clients to collaboratively train a global model using their local datasets while reinforcing data privacy, but it is prone to poisoning attacks. Existing defense mechanisms assume that clients' data are independent and identically distributed (IID), making them ineffective in real-world applications where data are non-IID. This paper presents FL-CLEANER, the first defense capable of filtering both byzantine and backdoor attackers' model updates in a non-IID FL environment. The originality of FL-CLEANER is twofold. First, it relies on a client confidence score derived from the reconstruction errors of each client's model activation maps for a given trigger set, with reconstruction errors obtained by means of a Conditional Variational Autoencoder trained according to a novel server-side strategy. Second, it uses an original ad-hoc trust propagation algorithm we propose. Based on previous client scores, it allows building a cluster of benign clients while flagging potential attackers. Experimental results on the datasets MNIST and FashionMNIST demonstrate the efficiency of FL-CLEANER against Byzantine attackers as well as to some state-of-the-art backdoors in non-IID scenarios; it achieves a close-to-zero (<1\%) benign client misclassification rate, even in the absence of an attack, and achieves strong performance compared to state of the art defenses.
\end{abstract}




\maketitle

%-----------------------------------------------------------------------
\section{Introduction}
%-------------------------------------------------------------------------------

In recent years, Federated Learning (FL) has attracted considerable interest as a privacy-preserving machine learning paradigm \cite{matta2023federated,ji2024emerging,guan2024federated} in several domains, such as healthcare \cite{lansari2023federated}, transportation \cite{lansari2023federated}, and personal devices \cite{smartkeyboardgoogle}. In a typical FL scenario, several clients collectively contribute to training a global model. At first, a central server initiates the global model and sends it to the clients so that they can train it locally on their private data. Model updates, in the form of weights, are sent back to the server which aggregates them into a new global model. These two steps are repeated until the global model converges. As clients' data remain stored locally, privacy is ensured \cite{mcmahan2016federated}.
However, even though FL systems preserve privacy, their decentralized nature makes them vulnerable to poisoning attacks, where one or several malicious clients try to corrupt or prevent the training session by sending malicious updates \cite{xia2023poisoning}. Some defenses try to mitigate the impact of these attacks using secure aggregation algorithms with the idea of reducing the impact of bad updates \cite{mcmahan2017communication,blanchard2017byzantine,guerraoui2018hidden}. Nevertheless, these solutions remain limited. They are, for instance, not resistant to attacks able to destroy the global model after just one successful FL round of attack, known as Byzantine attacks \cite{chen2022attacks}. 
This has led to the proposal of FL defense systems devoted to the detection of malicious updates before the aggregation stage. Solutions proposed to date \cite{gu2021detecting,chelli2023fedguard,castillo_fledge_2023,rieger_deepsight_2022,bellafqira_fedcam_2024} have been proven to be Byzantine attack-robust under the hypothesis of Independent and Identically Distributed (IID) data, i.e., when training data is homogeneously distributed among clients in terms of the number of samples and class balance. However, such systems do not perform well in non-IID contexts \cite{allouah_robust_2023}, which are more likely encountered in real-life scenarios. In the following, we introduce these solutions pointing out their limitations before then presenting our contribution: FL-CLEANER, the first FL defense system adapted to non-IID contexts. Before concluding, we will also experimentally show that FL-CLEANER: i) effectively blocks all byzantine attacks in a non-IID setting, all while having near zero false positives; ii) mitigates the effects of backdoor attacks.

\begin{figure}[t]
  \centering
  \includegraphics[width=\linewidth]{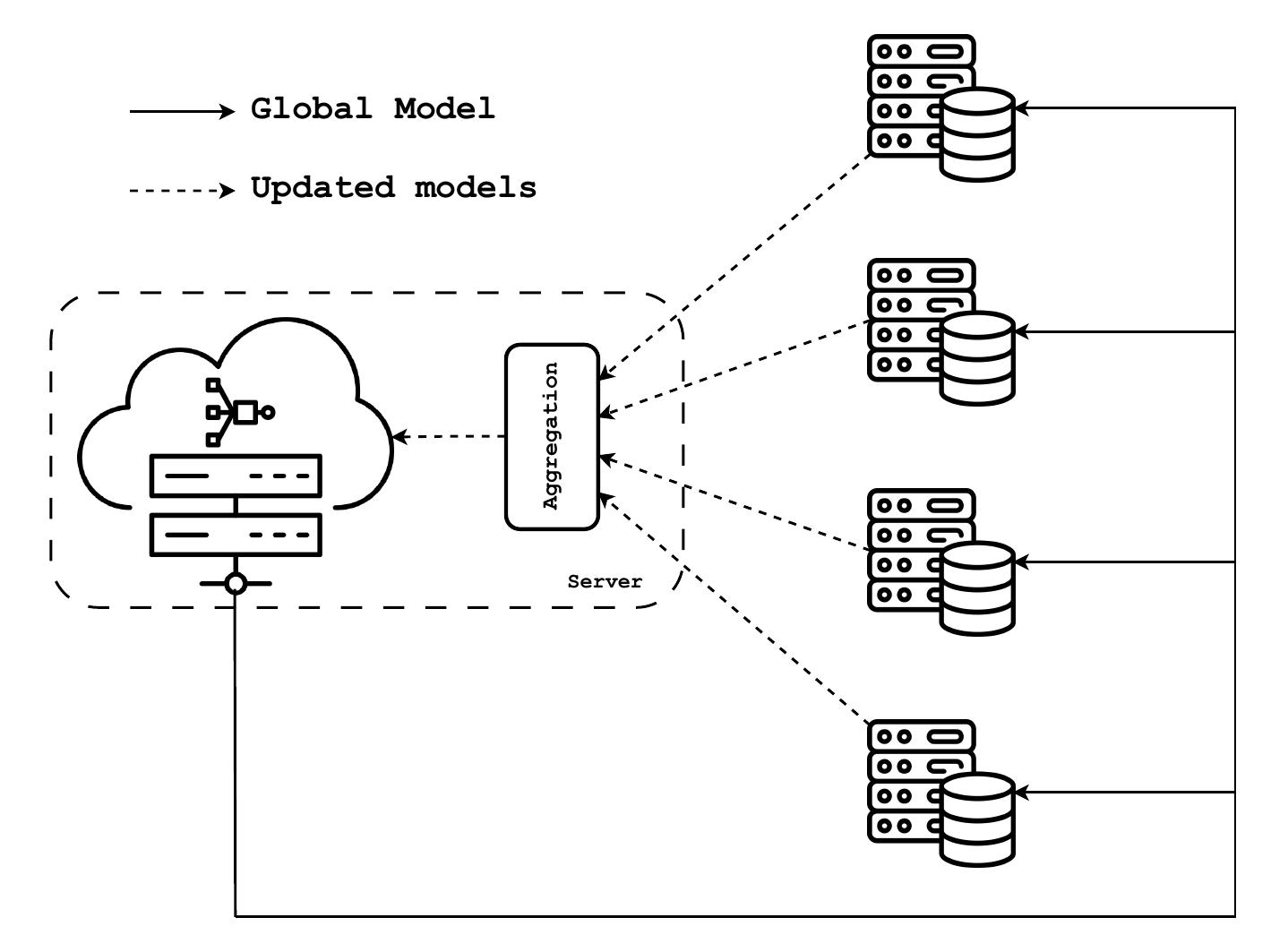}
  \caption{A common online FL scenario with a central server}
  \label{fl}
\end{figure}

\subsection{Related Works}
\label{subsec:relatedworks}
As stated above, the first defenses against malicious clients' updates were the development of robust aggregation algorithms. FedAvg \cite{mcmahan2017communication} is one of the earliest and most commonly used aggregation methods in FL. It consists of using a weighted average of the updates sent by clients to actualize the global model. Clients' dataset sizes are used as weights. Doing so dilutes the effect of attackers when they are few in number but is not sufficient when their number increases. It is also not robust to FL attacks that amplify updates to dominate the global average \cite{severi2022networklevel}, and ineffective in non-IID scenarios especially when an attacker has a larger dataset compared to benign clients. GeoMed \cite{guerraoui2018hidden} uses the geometric median of clients' updates as a global update. KRUM \cite{blanchard2017byzantine} uses one of the local updates to generate a global model update. The chosen local update minimizes the sum of distances to its nearest neighbors. Other aggregation schemes have been proposed since \cite{li2020federated,karimireddy2020scaffold,acar2021federated,gao2022feddc}, but they focus on training performance rather than security. All of these methods have been shown to be ineffective under Byzantine attacks \cite{gu2021detecting}.

To better respond to attacks, "detect and remove" defense systems have been recently proposed. They aim at excluding malicious updates before the aggregation step. One first class of methods regroups systems which rely on anomaly detection techniques based on autoencoders. Li \textit{et al.} \cite{li2020learning} were the first to propose such a defense. It trains a copy of the global model on the test dataset on the server side, while training at the same time a variational autoencoder (VAE) on surrogate vectors (randomly sampled weights from the model) at each epoch. Although this solution performs better than aggregation-based defenses under Byzantine attacks, it requires training a VAE on an entire server-side dataset to achieve good results. This is quite unrealistic in real-life scenarios, as the server most likely won't have access to a significant training dataset. FedCVAE \cite{gu2021detecting} later proposed an improvement using an untrained conditional variational autoencoder (CVAE). When the central server receives updates from the clients, the defense system computes the geometric median (GM) of surrogate vectors from each update. Then it normalizes these vectors by subtracting their GM value. These normalized vectors are then inputted into the CVAE, conditioned on the current round number in the FL training. As the CVAE is not pre-trained, only inputs near zero yield reconstruction errors close to zero, and vice versa. FedCVAE shows some defense results in a non-IID scenario but only against attacks exaggerated in magnitude and thus fairly easy to detect. They also provide results on simple backdoor attacks. More recently, FedCAM \cite{bellafqira_fedcam_2024}, a CVAE anomaly detection-based approach, achieves the latest state-of-the-art results against Byzantine attacks in IID scenarios. Rather than using model weights, it works with their activation maps for a given trigger set on which a CVAE is trained with GeoMed as a regularization. However, FedCAM only works in IID scenarios, as we will see in Section \ref{sec:experimental}.

A third class of methods, as introduced by FedGuard \cite{chelli2023fedguard}, aims at exploiting clients' training data for detection without disclosing it. FedGuard \cite{chelli2023fedguard} asks each client to train a local CVAE on its own dataset in addition to the main task and to send to the server both its model updates and the decoder part of its CVAE. The server then uses all client decoders to generate sample data to evaluate each received local model update's performance. Then for the aggregation step, it only keeps models that achieve a certain accuracy. This solves the reliance on an auxiliary dataset for CVAE training. However, it incurs significant computing overhead and demands heavy client resources. In addition, this approach also depends on the clients having enough data to train a generative model. This is achievable in an IID scenario with a small number of clients but unlikely in a non-IID scenario where datasets vary in size and content. More importantly, this solution opens up clients to data leakage from the generative decoders through, for instance, membership inference attacks \cite{shokri2016membership}.

A last class corresponds to clustering-based approaches. They have shown promising results on anomaly detection without needing autoencoders. Among recent ones, we can cite FLEDGE \cite{castillo_fledge_2023} which used a combination of cosine similarity and Kernel Density Estimation to filter out attacks. It focuses on backdoor attacks but provides results against Byzantine attacks but only a fraction of their data is non-IID. 
Finally, Sharma et al. \cite{sharma2025detectionofmalicious} recently proposed a defense based on graph representation combined with Graph Attention Networks. Beyond the fact they require pre-training a model on both attacker and benign client examples, its authors provide only results for one byzantine attack, being sign-flipping attack. 

In addition to these defenses, several others exist that specifically target backdoor attacks \cite{NGUYEN2024107166}. Defenses such as FoolsGold \cite{fung2020mitigatingsybilsfederatedlearning}, DeepSight \cite{rieger_deepsight_2022}, FLAME \cite{nguyen2022flame} and FLDetector \cite{zhang2022fldetectordefendingfederatedlearning}rely on filtering out attackers by finding clients that present similar behavior. Other propositions try to remove the effect of the backdoor after the aggregation step, such as \cite{wu2021mitigatingbackdoorattacksfederated} and \cite{wu2022federatedunlearningknowledgedistillation}. However, these propositions have not been tested or proven effective against byzantine attacks. 

\subsection{Contribution}
In this work, we propose FL-CLEANER, an extension of FedCAM, introduced in \cite{bellafqira_fedcam_2024}, which was only devoted to byzantine attack detection in IID settings. Similarly to FedCAM, FL-CLEANER analyzes the GeoMed normalized CVAE reconstruction errors of clients' activation maps for given inputs of a trigger set. Clients with high reconstruction errors are rejected based on an anomaly detection strategy. While FedCAM achieves very good byzantine robustness compared to concurrent methods in IID scenarios, it fails to achieve good results in the context of non-IID data, as we will see in Section \ref{sec:experimental}. Two main issues cause this failure. First, the CVAE training method proposed in FedCAM has poor generalization in non-IID contexts, as client updates are more heterogeneous. Specifically, clients with the most unique data often fall out of distribution for CVAE and end up with similar error scores to attackers. To overcome this issue, FL-CLEANER employs a different CVAE training process that accounts for the temporal evolution of FL training with warmup epochs and better captures the statistical properties of benign updates using an annealing technique and layer selection. This ensures that all benign client activation maps, even outliers, are correctly reconstructed. Second, FedCAM relies on the mean of reconstruction errors as a dynamic error threshold. This approach can allow the selection of some attackers' updates for the aggregation step if the average is high enough. It is important to note that retaining even one Byzantine attacker in a round is sufficient to destroy the global model. Moreover, this strategy eliminates around half of the participants in each round. In the non-IID case where clients are less redundant compared to IID scenarios, this impacts the convergence and accuracy of the global model. To address this problem, FL-CLEANER introduces a clustering algorithm that iteratively aggregates and concentrates benign candidates into a trusted cluster, enabling a more refined selection process.

The rest of this paper is organized as follows. Section \ref{sec:background} provides an overview of FL technical background, FL attacks, and useful components of our solution. Section \ref{sec:model} presents the federated context we considered and its threat model. Section \ref{sec:overview} introduces our proposal, FL-CLEANER, followed by Section \ref{sec:experimental}, which details our benchmark and provides an analysis of the experimental results. Finally, we conclude the paper in Section \ref{sec:clc}.

%-------------------------------------------------------------------------------
\section{Background}
%-------------------------------------------------------------------------------
\label{sec:background}
\subsection{Federated Learning (FL)}
\subsubsection{A Common FL Scenario}
In this work, we considered an FL setting where $N$ clients $\{C_k\}_{k=1..N}$ with their datasets $\{D_k\}_{k=1..N}$ participate in a training session orchestrated by a central server $S$ to train a global model $M_G$. During FL round $t$, $S$ scatters the current global model $M_G^t$ to a subset of $U$ randomly selected clients. Each client updates the received model on its local dataset and sends back the updated model to $S$. The server then uses an aggregation algorithm such as FedAvg \cite{mcmahan2017communication} to get an updated version of the global model $M_G^{t+1}$ from the clients' updated models $\{M_u^t\}_{u=1..U}$. The above training steps are then repeated until convergence or a stopping criterion is met.

\subsubsection{Non-IID Data in Federated Learning}
One major statistical challenge in FL is the so-called Non-Independent and Identically-Distributed (Non-IID) hypothesis \cite{kairouz2021advances,DBLP:journals/corr/abs-1806-00582}. Indeed, in most real-life applications \cite{el2024secure}, local datasets are client-specific, and data collection processes are often inconsistent across clients, resulting in data heterogeneity.
In the literature, four main heterogeneity types are identified \cite{ye2023heterogeneousfederatedlearningstateoftheart}:

\begin{itemize}
\item \textbf{Label skew:} Where class labels are distributed differently among clients.
\item \textbf{Feature skew:} Where clients have differently-distributed features for the same class.
\item \textbf{Quality skew:} Where data quality is inconsistent across clients.
\item \textbf{Quantity skew:} Where the amount of local data may be unbalanced across clients.
\end{itemize}

Non-IID data hypothesis is considered a challenge for FL defenses for two main reasons. First, a well known problem in FL literature \cite{zhao2018federated,kairouz2021advances} is that it is harder to make the global model converge in this setting, since clients with different data often send contradicting updates. From a defense point of view, this means that it is a risk to misclassify honest clients, requiring precise clients' udpates filtering. Second, defenses that rely on detecting attackers often look for statistical anomalies to spot these attackers. This is more easy in a IID data context where clients share similar data distribution and thus present similar statistical properties. With non-IID data, clients naturally have different characteristic, and it becomes more difficult to differentiate attackers from natural outliers.

In practice, feature and quality skew are intrinsic properties of data and are hard to simulate experimentally if the dataset does not provide a natural partition. So, most works that address Non-IID FL use a combination of label and quantity skews by choosing a distribution of data over clients that ensures heterogeneity.
\subsubsection{Attacks on Federated Learning}
Federated Learning is known to be vulnerable to several attacks \cite{kairouz2021advances}. These attacks operate at training time and involve an adversary that compromises one or several clients and tries to use them to disrupt the global model. Two main approaches can be distinguished: "model poisoning", where the attacker can directly manipulate the compromised clients' models and send malicious updates to the server; and "data poisoning", where the attacker indirectly manipulates learning by corrupting the clients' local datasets. Based on the attacker's objective, these attacks can further be divided into two sub-classes:
\begin{itemize}
\item \textbf{Untargeted poisoning attacks} - where the attacker's goal is to drop the global model accuracy on any future inputs by preventing its convergence. It is generally very hard for a machine learning model to recover from a bad convergence, and one attack in an unprotected FL round is often sufficient for an untargeted poisoning attack to succeed \cite{blanchard2017byzantine}. Such attacks are also referred to as "Byzantine attacks".
\item \textbf{Targeted poisoning attacks} - also called "Backdoor attacks", where the attacker aims at preserving the accuracy of the model on its main task while simultaneously training it on a second malicious task, introducing a backdoor. The most common form of backdoor attacks involves inserting a trigger in the training dataset and training the model to output a specific response when the trigger data are given as an input. \cite{han2024badsflbackdoorattackscaffold}. Through the control of just some of the FL clients, the attacker may have insufficiently representative data because of the Non-IID context  \cite{liu2022technical}, making the backdoor updates easily overwritten by future training rounds, as shown in \cite{liu2022technical}
\end{itemize}

\subsection{Conditional Variational Autoencoders (CVAE)}

\begin{figure*}[!t]
  \centering
\includegraphics[width=0.65\linewidth]{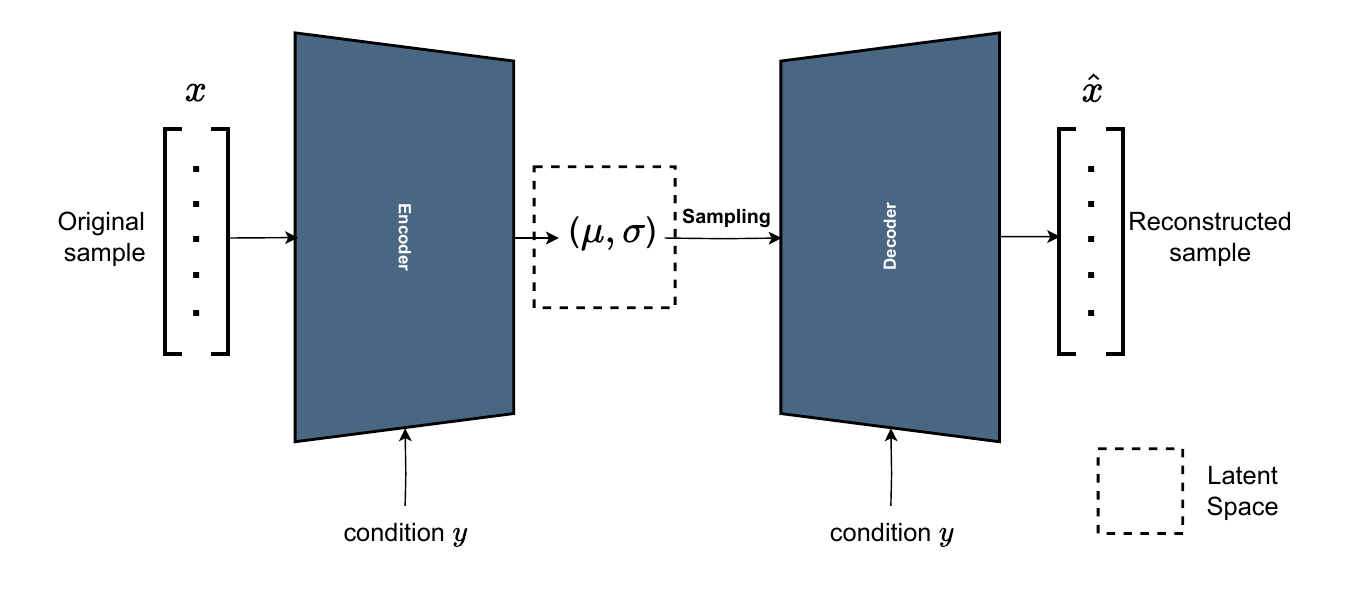}
\caption{Architecture of a Conditional Variational Autoencoder (CVAE), where $x$ and $\hat{x}$ are the original and the reconstructed samples, respectively. $y$ is the condition, and $\mu$ and $\sigma$ are trainable parameters of the CVAE latent space.}
  \label{fig:cvae_ex}
\end{figure*}

Autoencoders (AEs) \cite{hinton2006reducing} are a class of unsupervised generative neural networks that aim to reconstruct a given input from a latent representation encoding. AEs are made of: i) an encoder $f$ that maps input vectors from the initial feature space to a lower-dimensionality representation space called the latent space; and ii) a decoder that retrieves input vectors from their latent representations. The objective of an autoencoder is to minimize the reconstruction error (RE) between the initial and reconstructed vectors. They have been historically been used for dimensionality reduction \cite{Belkacemi_2023_deminsionality_reduction,hinton2006reducing}, model pre-training \cite{erhan2010whydoes,bengio2007greedy}, and generative tasks \cite{yan2016attribute2imageconditionalimagegeneration}. An important property of AEs is that, when applied to out-of-distribution vectors that deviate from the original training distribution, these deviations are amplified in the latent space, and much higher reconstruction errors are observed. This makes AEs considered as a valuable tool for anomaly detection in a wide range of applications \cite{pang2021deep}.

Autoencoders having a fixed-dimensional latent space show some limitations for datasets with varying complexities. This has motivated the introduction of Variational Autoencoders (VAEs) \cite{kingma2013auto} with a variational latent space based on Bayesian inference methods \cite{doersch2016tutorial}. Conditional-VAEs (CVAEs) \cite{pol2019anomaly} are an extension of VAEs that leverages conditional probabilities to guide the encoding and decoding process following a provided condition. In practice, a CVAE is a VAE with a condition vector $y$ as additional input for both the encoder and decoder (see Fig. \ref{fig:cvae_ex}). CVAEs have been extensively used for a number of applications, including anomaly detection and detecting attackers in federated learning \cite{gu2021detecting,bellafqira_fedcam_2024,chelli2023fedguard}.

\subsection{GeoMed and Activation Maps}
In a Euclidean space, the geometric median (GeoMed) of a set of points is a point minimizing the $L_2$ distance to all the points in the set. GeoMed is a robust approximation of the centroid of a given set of points under the presence of outliers \cite{guerraoui2018hidden}. This has led to its extensive use in Byzantine-robust federated learning as a robust aggregation method \cite{chen2017distributed} or to detect attackers \cite{gu2021detecting}.

In a typical deep learning model, data are processed through feed-forward layers. Each hidden layer takes in the previous layer's output, applies a linear transformation parameterized by a set of weights (e.g., a convolution or weighted sum), and then applies a nonlinearity function. The output of the layer's weighted transformation is referred to as the activations of that layer. "Activation maps", or the set of activations extracted from a model for a given input, are characteristic of a model and the way it processes data. They have been used for applications such as explainability \cite{selvaraju2017grad-cam} and model watermarking \cite{darvish2019deepsigns,bellafqira2022diction,zheng2024high}. It has been observed that in many real-life applications, the activation maps of a model tend to follow an intrinsic distribution \cite{darvish2019deepsigns,patel2015probabilistictheorydeeplearning}.

\section{Threat Model and Defense Requirements}
%-------------------------------------------------------------------------------
\label{sec:model}
In this section, we present the threat model and security requirements considered in this work. In our FL scenario, we assume that the clients never share their datasets and only communicate their updated models to the server. Regarding the server, it has access to a sample of the testing set and the updated models received at each round (i.e., the server can inspect the inner layers and weights of the models clients send back after training).

\subsection{Threat Model}
\label{subsec:threatmodel}
\subsubsection{Adversary's Objective}
In this work, we consider a "Byzantine/backdoor" adversary $A$, who wishes to carry out Byzantine/backdoor attack during the training phase of an FL model. To do so, $A$ controls a number of the FL clients who submit poisoned model updates to $S$ during the FL process with the objective of disturbing the convergence of the global model. Formally,  for Byzantine attacks, $A$ attempts to fulfill the following objectives:
\begin{itemize}
\item \textbf{O1}: If the attack succeeds, the global model should have bad accuracy on test data.
\item \textbf{O2}: The attacked model should not be able to recover from the accuracy drop by pursuing further training in the absence of the attacker.
\end{itemize}
And, for a backdoor to be successful : 
\begin{itemize}
    \item \textbf{O3}: Achieve a good attacker success rate (ASR), which corresponds to the ratio of test samples with the backdoor trigger that the model successfully classifies in the target class.
\end{itemize}

\subsubsection{Adversary's Capabilities}
As in \cite{bagdasaryan2020backdoor,rieger_deepsight_2022,fang2020local,chelli2023fedguard,gu2021detecting,castillo_fledge_2023}, we consider a strong adversary who has full control over $m < N/2$ "malicious" clients; $N$ being the total number of clients. $A$ has full access to datasets, models, and communication channels of these clients. In particular, $A$ can conduct training using local datasets when clients are selected by the server, freely modify the weights of the resulting model updates, and send back modified updates to the server $S$ for aggregation. In accordance with Shannon's Maxim, $A$ also has full knowledge of the aggregation method used by $S$ and of the defense this one deploys but cannot access or alter its parameters. They also have no access to other benign clients' updates or datasets.

\subsection{Security Requirements}
\label{subsec:securityrequirements}
To effectively defend against a Byzantine or backdoor attacker, it is required that a filtering-based defense system $D$ satisfies the following requirements \cite{rieger_deepsight_2022}: 
\begin{itemize}
\item \textbf{R1: Block all malicious byzantine clients}. As discussed above, the Byzantine adversary is able to break the model given one single successful round of attack. As a result, in order to be robust to such an adversary, $D$ must detect all of these malicious client updates at any round.
\item \textbf{R2: Do not disrupt training}. $D$ should not negatively impact the training. In other words, $D$ must have high precision and not misclassify benign clients. It must also preserve the test accuracy achieved in the baseline no-attack and no-defense scenario.
\item \textbf{R3}: \textbf{Lower backdoor attacker success rate}. In the presence of the defense, the backdoor accuracy measured on the test set should be significantly lower.
\end{itemize}
We will see in the sequel that our proposal fulfills these three security requirements.

%-------------------------------------------------------------------------------
\section{Proposed defense: FL-CLEANER}
%-------------------------------------------------------------------------------
\label{sec:overview}

\subsection{FL-CLEANER  Basic Principles}

 The idea behind FL-CLEANER is to analyze the normalized activation maps (NAMs), obtained by subtracting the GeoMed value from activation maps of one or several layers of $M_G$ for some given inputs referred as a trigger set $T_s = (X_T, Y_T)$, where $X_T$ and $Y_T$ are the set of samples and their labels, respectively. We assume that NAMs obtained with malicious model updates will correspond to outliers that can be identified by means of a CVAE and GeoMed based anomaly detection strategy. 
 
In a simplified view, before the FL process begins, FL-CLEANER :
 \begin{enumerate}
     \item Makes a copy of the initial global model. 
     \item Trains this copy locally for several epochs on a "trigger set" that is a very small dataset. 
     \item Extracts NAMs associated with the trigger set from different epochs of training
     \item Trains a CVAE to reconstruct these NAMs, using their corresponding labels $Y_T$ as a condition. The training process is detailed in \ref{subsec:training}.
 \end{enumerate}
By doing so, the CVAE learns the latent distribution of NAMs associated with each class. 

Once trained, $S$ uses CVAE at each epoch to reconstruct NAMs corresponding to the trigger set for each client updated model $S$ received. Malicious client NAMs, which fall out of the learned distribution, result in high reconstruction errors. To discriminate malicious NAMs from benign ones, FL-CLEANER makes use of a clustering procedure relying on an ad-hoc trust propagation algorithm. We present and discuss it in subsection \ref{subsec:trust}.

 \medskip
To go in further details and provide a formal description of FL-CLEANER, let us consider a FL scenario with a classification task and $N$ clients $\{C_k\}_{k=1..N}$. FL-CLEANER operates according to the following steps: 

\begin{enumerate}
    \item \textbf{FL and FL-CLEANER Initialization ($t=0$)}  
    \begin{enumerate}
        \item $S$ initializes the global FL model $M_G^0$ in a classic way assigning, for instance, random values to its weights. 
        \item Starting from $M_G^0$, $S$ trains its own version of the model: $M_G^{'0}$, on its trigger set $T_s = (X_T, Y_T)$, where $X_T$ and $Y_T$ are the set of samples and their corresponding labels. 
         \item $S$ then trains a CVAE model. To do so, it first extracts the activation maps $A_S$ of one or more layers from $M_G^{'0}$ feeding it with the trigger set $T_s$. Then, it normalizes them using GeoMed such as: $\hat{A}_S= \sigma(A_S-\text{GeoMed})$, with $\sigma$ is the sigmoid function. $S$ finally trains CVAE on $\hat{A}_S$.
        \item $S$ initiates the first FL training round by sending $M_G^0$ to $U$ randomly selected clients ($U < N$),.          
    \end{enumerate}

    \medskip
    
    \item \textbf{ Round $t\geq1$ of the FL training process}  
    \begin{enumerate} 
        \item \textbf{Client training}
        \begin{enumerate}
        \item $S$ randomly selects $U$ clients $(U<K)$ to participate in the round $t$ and sends them the current global model $M_G^{t-1}$.
        \item Each client $C_u$ trains $M_G^{t-1}$ on its local dataset $D_u$ for $n$ local epochs. The resulting updated model $M^t_u$ is next sent back to the server. 
        \end{enumerate}
        \item \textbf{Secure FL-CLEANER selection}
        \begin{enumerate}
        \item \textbf{Activation map inspection for each client $u$ :} $S$ extracts the activation maps $A_u^t$ obtained by feeding the trigger set $T_s$ into $M^t_u$. 
        \item \textbf{Scoring:} 
        \begin{itemize}
            \item $S$ computes $\text{GeoMed}(A_1^t,...,A_U^t)$, that is the GeoMed centroid of all activation maps. Then it normalizes the activation maps of each client: $\hat{A}_u^t= \sigma(A_u^t-\text{GeoMed})$.
            \item $S$ feeds its CVAE with $\hat{A}_u^t$ conditioned on $Y_t$ , the label, and assigns to each client a reconstruction error score $\epsilon_u^t=\overline{MSE}(\hat{A}_u^t,\Tilde{A}_u^t)$ where $\Tilde{A}_u^t$ is the CVAE reconstruction of $\hat{A}_u^t$, and $MSE$ the mean square error function. 
        \end{itemize}
        \end{enumerate}
        \item \textbf{Filtering:} Given the pairs $(M_u^t,\epsilon_u^t)$, $S$ performs our trust propagation clustering (see next Section) to filter out clients that are considered malicious and discards their updates. 
        \item \textbf{FL Aggregation:} Finally, $S$ aggregates updates from the remaining clients using FedAvg to generate $M_G^t$. 
    \end{enumerate}
\end{enumerate}

As exposed, the performance of FL-CLEANER to discriminate malicious from benign updates rely on two main components: a CVAE model and a trust propagation clustering algorithm. We come back to how they work in the following sections. 

\subsection{CVAE training}
\label{subsec:training}
As stated above, the CVAE model is trained by the server $S$ considering a simulated benign model $M_G^{'0}$ based on the server trigger set $T_s$. In this work we assume that $T_s$ is of very small size compared to the client data. This ensures the real-life applicability of our solution as servers may not have access to the large datasets other methods in literature require. Moreover, servers may have some data of their own or access to trusted clients or can form such datasets from open-source data. The influence of the trigger set size and content is not included in our study due to space limitations. However, our experiments showed that it is possible to achieve good defense results against Byzantine attackers with as little as 250 images randomly sampled from the server test set and 500 images for backdoor attacks.

\begin{figure*}[!t]
  \centering
  \makebox[\linewidth]{
  \includegraphics[width=1.05\linewidth]{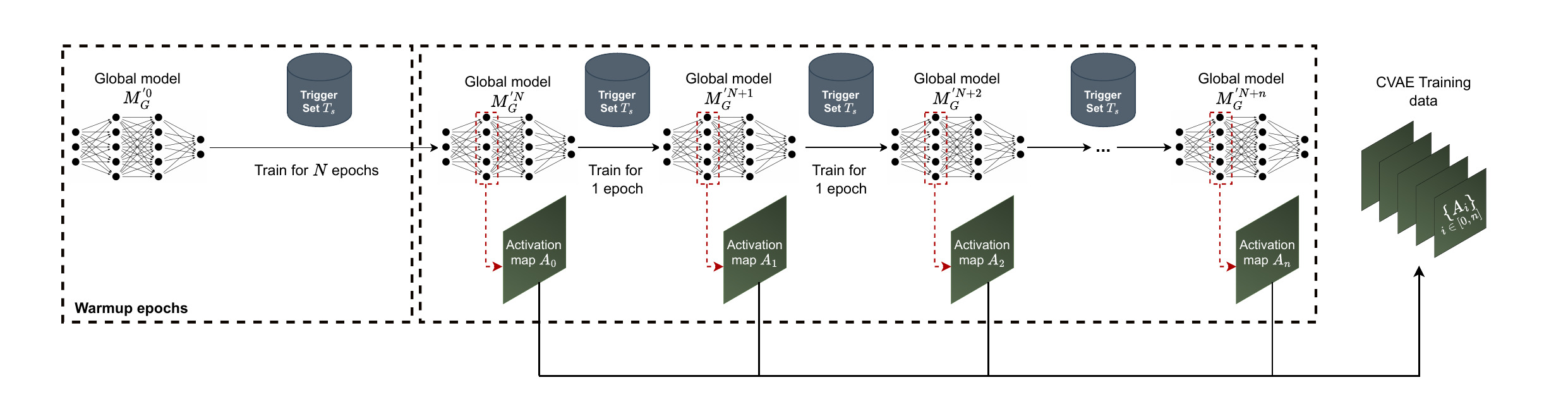}}
  \caption{Construction of the CVAE training set.}
  \label{cvae_train}
\end{figure*}

To augment the number of training samples available to the CVAE with a small dataset, the solution we adopted is to use the activation maps (AMs) computed at different rounds of the training of $M_G^{'0}$. AMs are more heavily affected by noise and vary a lot from one epoch to another at the very beginning of the training of $M_G^{'0}$. To account for this, as whown in in Fig. \ref{cvae_train}, the CVAE training set is constituted of activation maps collected after some warmup epochs of the training of $M_G^{'0}$. Additionnally, AMs are GeoMed normalized into NAMS at each epoch so as to better control AMs amplitude differences along the training of $M_G^{'0}$. Doing so also helps to take into account the temporal drift of $M_G^{'0}$ activation maps. Indeed, the distribution of the deviations from GeoMed recomputed at each iteration remains stable throught the epochs. That is not the case of the distribution of the AMs which drifts over time. We will see that even with a small trigger set, these technique allows to constitute a training set that well captures the distribution of benign NAMs, and ensures good CVAE generalization capabilities.

A second issue to consider is that CVAE training is sensitive to "posterior collapse" \cite{higgins2017betavae,bowman_generating_2016,dang2024vanilla},  causing it to converge to minima that ignore training inputs. Indeed, the traditional training of a CVAE, consists in maximizing the Evidence Lower Bound (ELBO) in the following loss:  
\vspace{-0.3cm}
\begin{multline}
    \mathcal{L}(x,y;\theta,\phi) = 
    -\text{ELBO}(x,y;\theta,\phi)\\
    = -(\mathbb{E}[\log p_{\theta}(x|z,y)] - \frac{1}{2}[\sigma_\phi(x)^2 + \mu_\phi(x)^2 - 1 - log(\sigma_\phi(x)^2)) \\
    = \text{MSE}(g_\theta(z,y),x) + D_{\text{KL}}(q_{\phi}(z|x) \| p(z))
\end{multline}
where: $g_\theta$ is the CVAE decoder; $q_\phi$ is the latent space distribution given by the CVAE encoder; $z$ is a latent representation sampled from this distribution; and, $(x,y)$ the CVAE input sample and its label; $D_{KL}$ is the Kullback-Leibler divergence measuring the divergence with a standard normal prior distribution $p$. As exposed, the first loss term minimizes the CVAE reconstruction error and the second  enforces a Gaussian structure for the latent space representation. In our context, NAMs give rise to numerically very small reconstruction error values making the optimizer to focus on minimizing the second term: the so called "posterior collapse" issue. To solve this issue, we add a regularization weight to the divergence term in the loss function \cite{higgins2017betavae}. In our work, we adopted a KLD annealing approach with an adaptive regularization weight $\beta^t$ as in \cite{bowman_generating_2016} 
\begin{equation}
    \mathcal{L}_{CVAE} = MSE + \beta^tD_{KL} 
\end{equation}
where, $t$ is the current training epoch. CVAE starts its training with $\beta^{0}=0$ Then, $\beta$ is gradually increased by a fixed increase to enforce the structure of the latent space. 

\subsection{Clustering using trust propagation}
\label{subsec:trust}
A commonplace practice in CVAE anomaly detection and FL defenses \cite{bellafqira_fedcam_2024,gu2021detecting,li2020learning} is to threshold error values to decide of an anomaly. Usually, this is the average reconstruction error of all clients \cite{bellafqira_fedcam_2024, gu2021detecting}. A result of this is that in the absence of an attacker almost half of the clients are wrongly flagged as malicious, even if they are benign. Well-adjusted static thresholds could allow for better filtering. In an IID scenario, such reference values could be implied from the error values of CVAE during training. However, in a Non-IID scenario, real client reconstruction error values cannot be deduced from the sample training scenario. We opted for a simple clustering algorithm instead.

Our clustering algorithm automatically outputs either one or two clusters depending on whether malicious updates are present. Notice that deciding the number of clusters based on data properties is an active problem in cluster analysis \cite{ezugwu2022survey}. Indeed, most algorithms require fixing the number of clusters beforehand as a user-defined parameter \cite{ezugwu2022survey} with  clustering performance that varies widely in terms of "parameter reliance", i.e.  parameter tolerance ranges. Our proposal has a single parameter that is interpretable (i.e. not rely on guessing) and fixes the number of clusters automatically. 

\begin{figure*}[!t]
  \centering
  \includegraphics[width=0.8\linewidth]{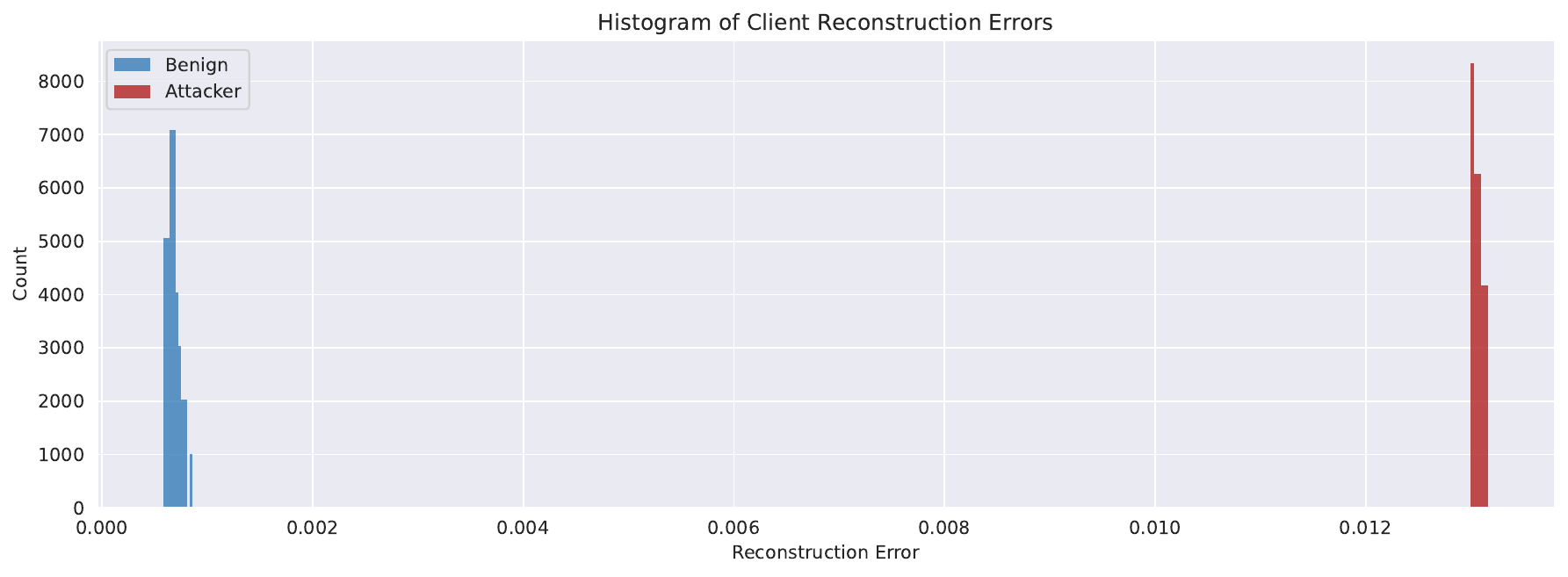}
  \caption{Histogram of reconstruction error values measured for benign clients and attackers performing a Sign Flipping attack on MNIST, considering a trigger set $T_s$ of 250 samples.}
  \label{fig:recon_errors}
\end{figure*}

The basic idea of our solution relies on the fact that for benign client's activation maps the measured reconstruction error is a residual from the randomness and uncertainty within CVAE and that the reconstructed vector in the case of an out-of-distribution attacker NAMs is an arbitrary response akin to measuring error to a random vector. Assuming that the CVAE has properly fitted benign NAMs distribution, expected reconstruction errors of benign NAMs are of a lower order of magnitude than for anomalous NAMs. In other words: in the reconstruction error space, there is an identifiable relative error gap between the two, as exemplified in Figure \ref{fig:recon_errors} in the case of a sign flipping attack (sign inversion of mdoel weights).

Based on this, we propose a "trust propagation" algorithm the principle of which consists in sequentially retrieving all benign clients participating in training. First, it sorts the client reconstruction errors, and selects as the core of the benign client cluster the one with the lowest error score. Then, in the reconstruction error space, if a client is within a certain distance $\delta$ of the last accepted client, it is added to the group. The iterations stop when the distance $d$ of the next client from the current client is such as $d>\delta$. This gap between error values corresponds to the frontier between attackers and benigns. All the remaining clients constitute the attacker cluster, and are blocked. Our algorithm depends on one single parameter $\delta$  which can be dynamically computed  as a proportion of the size of the error space for the current round, such as: 
\begin{equation}
\delta=\lambda(max(\epsilon_u^t)-min(\epsilon_u^t))_{0<u<U}
\end{equation}
Where $\lambda \in [0,1]$ is a hyper-parameter fixed in a data-agnostic manner at the start of the defense system. The full pseudo-code for the trust propagation algorithm is detailed in Alg. \ref{alg:cap}.

By sequentially building the trusted clients cluster rather than fixing a threshold, our detection handles attacks but also normal scenarios where the highest error clients are still benign. We observed that benign clients are very rarely misclassified with our method, excluding less than 1\% of benign clients in the worst case (see \ref{subsec:results}). 

\begin{algorithm}[H]
\caption{Trust propagation}\label{alg:cap}
\begin{algorithmic}
\REQUIRE $\delta$, clients\_sorted, sorted\_re
\STATE  $\text{benign\_clients} \gets \text{clients\_sorted[0]}$
\STATE  $\text{i} \gets \text{0}$
\STATE  $\text{i\_{max}} \gets \text{len(clients\_sorted)-1}$
\WHILE{($\text{i} \neq \text{i\_{max}}$) \textbf{and} \text{(sorted\_re[i+1]} $\leq$ \text{sorted\_re[i] +} $\delta$)}
    \STATE  $\text{benign\_clients.append(clients\_sorted[i+1])}$
    \STATE  $\text{i} \gets \text{i + 1}$
\ENDWHILE
\STATE  \RETURN $\text{benign\_clients}$
\end{algorithmic}
\end{algorithm}

%-------------------------------------------------------------------------------
\section{Evaluation}
%-------------------------------------------------------------------------------
\label{sec:experimental}
\subsection{Experimental setup}
\subsubsection{FL setting}
We evaluated FL-CLEANER on Non-IID image classification tasks using two public datasets : MNIST \cite{deng2012mnist} and FashionMNIST \cite{xiao2017fashionmnistnovelimagedataset}. In the following, FL is orchestrated by a server that applies FL-CLEANER as defense $D$ before each aggregation step with 100 total clients. 50\% of clients are randomly selected to participate in training at each round, and we use FedAVG \cite{mcmahan2017communication} to aggregate model updates. As global model $M_G$, we use a standard CNN architecture that achieves good results on both datasets in the absence of attackers. In all our experiments, we set 30\% of the client population to be attackers, in accordance with previous works in literature \cite{bellafqira_fedcam_2024}.

\subsubsection{Non-IID Data distribution}

\begin{figure*}[!t]
  \centering
  \includegraphics[width=0.7\linewidth]{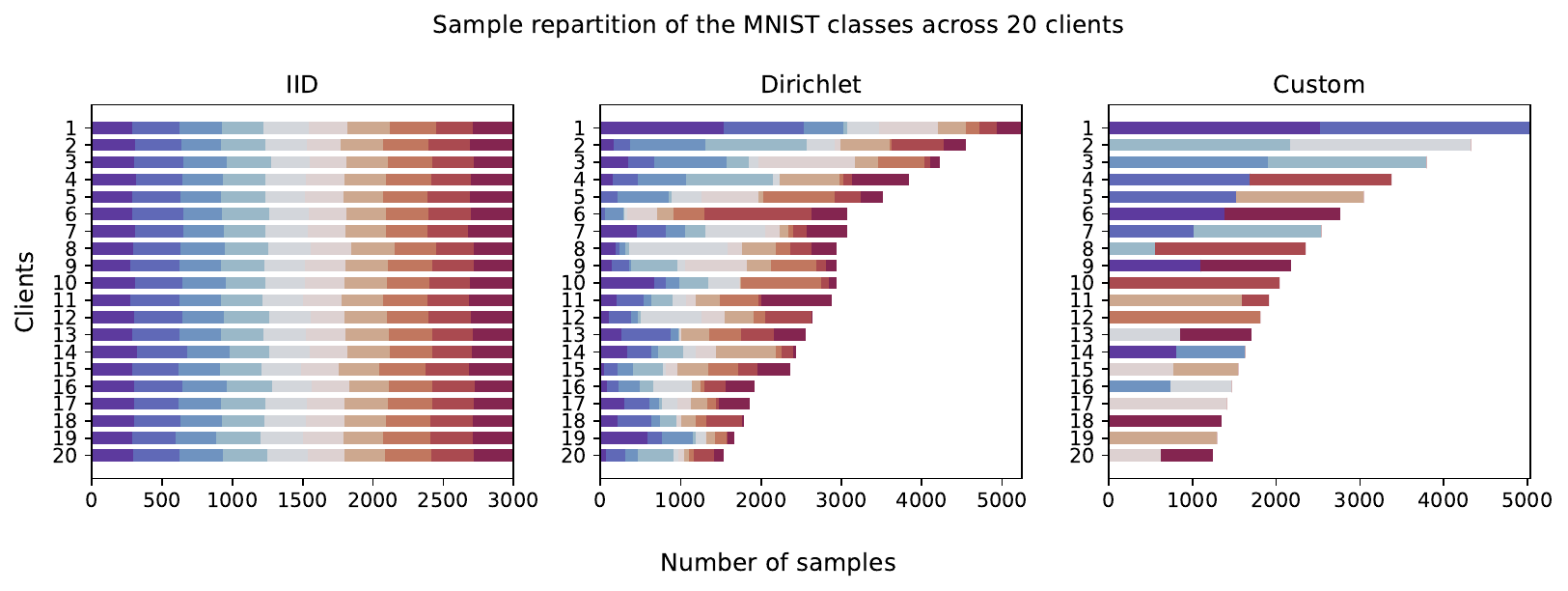}
  \caption{Examples of our Non-IID distributions of the MNIST training samples in a scenario with 20 total clients. Colors represent the image classes present in the data.}
  \label{fig:nIID_distribs}
\end{figure*}

To synthesize a Non-IID data distribution among clients, we distribute the samples of MNIST and FashionMNIST datasets to achieve:  
\begin{itemize}
    \item \textbf{Label skew} based on a latent Dirichlet distributions  as in  \cite{hsu2019measuringeffectsnonidenticaldata} and \cite{yurochkin2019bayesiannonparametricfederatedlearning}  - we synthesize a population of $C$ clients by drawing a probability vector $\mathbf{q_n} \sim D_C(\alpha)$ for each class $n\in[1,N]$. Each client is allocated a portion $q_{n,c}$ of the total samples in class $n$. We set $\alpha=1$ \cite{hsu2019measuringeffectsnonidenticaldata}.
    \item \textbf{Quantity skew} based on a custom distribution we propose where clients are only allowed to hold samples from two classes - we enforce the Non-IID condition originally proposed in \cite{mcmahan2017communication} where each client only possesses samples from two classes. This is achieved by sampling the full datasets using an inverse law with a scale factor of $\alpha$ and an offset of $\gamma$. For the $c^{th}$ client, we draw $D_s(c) = \lfloor \frac{\alpha}{c+r} \rfloor + \gamma$ samples without replacement from two arbitrarily selected classes. $r$ ensures $D_s$ doesn't outnumber class samples for small $c$ values. We set $\alpha = 2000$, $\gamma = 20$, and $r = 2$. 
   
\end{itemize}
Figure \ref{fig:nIID_distribs} shows an example of such distributions for the MNIST dataset.

\subsubsection{FL-CLEANER defense settings}
We use a CVAE conditioned on class labels. Its encoder and decoder are two-layer MLPs with a hidden layer size of 100 neurons. The number of CVAE training epochs has been fixed to 20. The KLD annealing weight was set to 0 with an increase of 0.5 after 10 epochs. The server $S$ trained its global model copy for 10 warmup epochs, then saved the activation maps from the next 10 epochs with a trigger set containing 250 samples extracted from the test set. We perform trust propagation with $\lambda=0.3$ (see Section \ref{subsec:trust}). In all our tests, we use all the layers of the model to extract activation maps. 

\subsection{Considered Byzantine attacks}

Four attacks inspired from literature were tested \cite{chen2022attacks,gu2021detecting,bellafqira_fedcam_2024}:

\begin{itemize}
\item \textbf{Sign-Flipping attacks}: Attackers reverse the signs of their model updates by multiplying them by a factor of $-\xi$. This is akin to reversing gradient descent and leading the model in the direction of errors. Let ${\omega_i}$ be the weights of the model update after local training. Attackers send back ${\omega'_i = -\xi\omega_i}$.
\item \textbf{Additive Noise attacks}: malicious clients add Gaussian noise to some of their model weights. Attackers send back ${\omega'_i = \omega_i + \epsilon_i}$, where $\epsilon_i \sim \mathcal{N}(0, \sigma)$.
\item \textbf{Same Value attacks}: malicious clients send a constant value $c$ as weight updates: ${\omega'_i = c}$.
    \item \textbf{Scaling attack}: malicious clients perform local stochastic gradient descent, then simply send back a scaled version of their updates, i.e., ${\omega'_i = a\omega}$ with $a>1$
\end{itemize}

The above parameters $\xi$, $\sigma$, $a$, and $c$  control the force or the amplitude of the corresponding attack striking the compromise in between attack efficiency and detectability. %, as the attacker updates have much higher numerical values than those of normal clients. 
For our tests to be nontrivial in terms of detection, we worked with attacks that send updates close to those of benign clients by fixing $\xi=1$, $\sigma=0.1$, $c=0.01$ and $a=10$. We will see that with no defense, these attacks are sufficient to collapse the global model, and that that they are also able to bypass detection by other defenses in literature (see Section \ref{subsec:results}). 

\subsection{Considered backdoor attacks}
For our experiment, we choose two state-of-the-art FL backdoor attacks, which have proven to be successful in non-IID scenarios: The distributed backdoor attack (DBA) \cite{yang2024distributedbackdoorattacksfederated} and Neurotoxin \cite{zhang2022neurotoxindurablebackdoorsfederated}. Let us recall that backdoor attacks work by adding a pattern (trigger) to training samples, changing their label to a target class, then training on these samples and sending back the corresponding model to the server. The two backdoors we tested work as follows :
\begin{itemize}
    \item \textbf{Distributed backdoor attacks (DBA)} \cite{yang2024distributedbackdoorattacksfederated} : Malicious clients work together to inject the trigger, by each adding a part of the pattern to their samples. When the resulting models are aggregated by the server, the global model associates the entire trigger with the target class
    \item \textbf{Neurotoxin} \cite{zhang2022neurotoxindurablebackdoorsfederated} : When training on the poisoned samples, the attackers only update the smallest $k\%$ components of the gradient, which are less likely to be overwritten during aggregation.
\end{itemize}
We test these attacks with a $10\times10$ pixels white square pattern for both MNIST and FashionMNIST, split into four $5\times 5$ squares for DBA. For Neurotoxin, we set $k=95\%$.

\subsection{Experimental results}
\label{subsec:results}

\begin{figure*}[!t]
  \centering
  \includegraphics[width=1\linewidth]{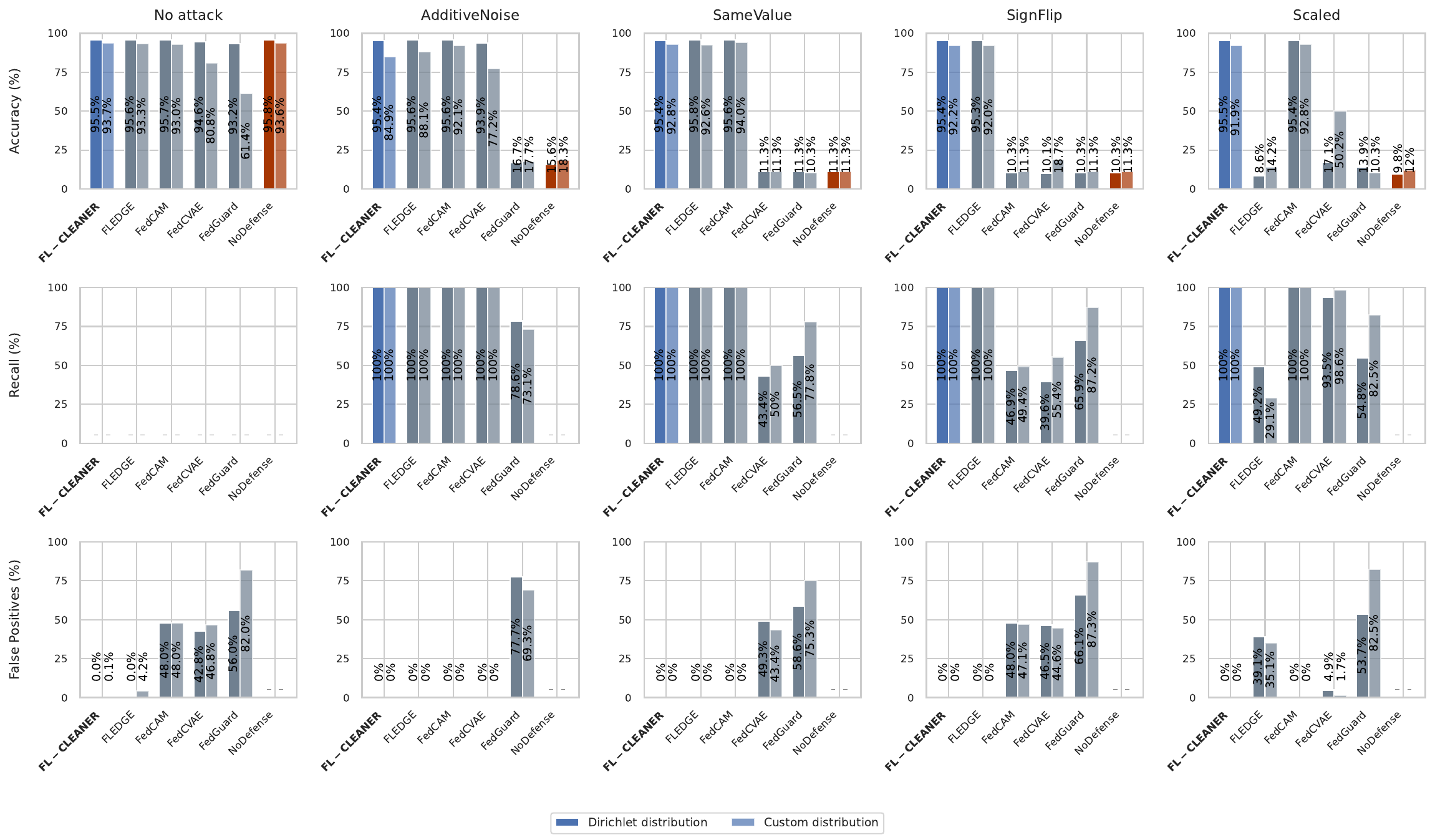}
  \caption{Defense Results on MNIST considering two non-IID distributions against Byzantine attacks: Additive noise, Same value, Sign Flipping and Scaled attacks of parameters $\xi=1$, $\sigma=0.1$, $c=0.01$ and $a=10$ respectively.}
  \label{fig:mnist}
\end{figure*}

\begin{figure*}[!t]
  \centering
  \includegraphics[width=1\linewidth]{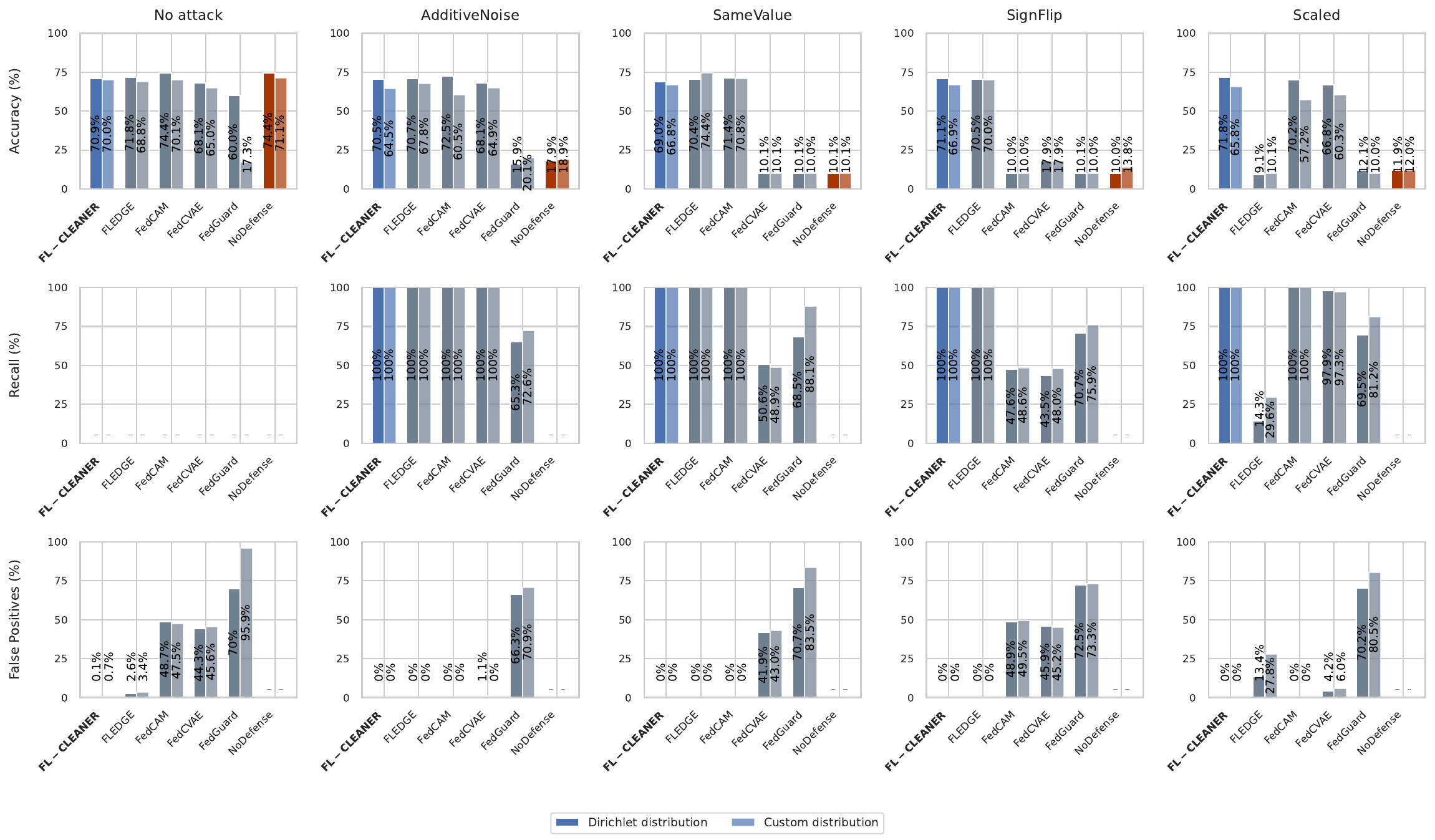}
  \caption{Defense Results on FashionMNIST considering two non-IID distributions against Byzantine attacks: Additive noise, Same value, Sign Flipping and Scaled attacks of parameters $\xi=1$, $\sigma=0.1$, $c=0.01$ and $a=10$ respectively.}
  \label{fig:fashionmnist}
\end{figure*}

For Byzantine attacks, we use three standard metrics to measure the effectiveness of each defense system: the global model's accuracy on the test dataset (\text{ACC}); the attacker detection recall (Recall) which refers to the percentage of attackers successfully blocked by the defense system; and, benign detection false positive (FPR) which represents the fraction of blocked benign clients. Figures \ref{fig:mnist} and \ref{fig:fashionmnist} report the results of testing on the two distributions of the MNIST and FashionMNIST datasets. For byzantine attacks, recall that a single attacker is sufficient to collapse the global model in just one FL round. Thus, a \text{Recall} value of 100\% is required in order the defense to be considered successful and reach the defense requirement \textbf{R1}  (see Section \ref{subsec:securityrequirements}). The defense also needs to not disrupt the training (\textbf{R2}), which means ACC should be as close as possible to the benchmark no defense and no attack accuracy, and \text{FPR} should be as close as possible to 0\%. As we can observe in Figures \ref{fig:mnist} and \ref{fig:fashionmnist}, compared to the most recent effective methods in literature (see Section \ref{subsec:relatedworks}), FL-CLEANER is the only defense system that achieves 100\% Recall against all types of Byzantine attacks in all tested scenarios. We can also see that when attackers are present, FL-CLEANER does not block any benign clients (i.e FPR=0\%). When attackers are absent, it misclassifies less than 1\% of benign clients. Additionally, across 20 tested scenarios, FL-CLEANER achieves the highest test accuracy in 10 scenarios and matches benchmark accuracy in all 20. Note that one reason why FL-CLEANER does not reach the highest accuracy achieved in the scenario with no-defense and no attack may come from the fact that in the experiments of FL-CLEANER only 70\% of the training data are used for training. The rest is used to simulate malicious clients for test purposes while in the scenario with no-defense and no attack, all data are used for training. Regardless, FL-CLEANER achieves 100\% attacker detection and 0\% or close miss-classification in each of these cases. 

It is important to note that other defenses were implemented according to the parameters provided by their authors. For FedCVAE for example, this meant using the whole test set, i.e, 10 000 images, to train the its CVAE. It can be seen that FL-CLEANER achieves better performance with a much smaller trigger set of 250 samples. 

\begin{table}[h!]
\centering
\setlength{\tabcolsep}{7pt} % adjust column spacing
\renewcommand{\arraystretch}{1.2} % adjust row spacing
\begin{tabular}{|l|c| c|c| c|}
\hline
\multirow{2}{*}{\textbf{Defense}} & \multicolumn{2}{c|}{\textbf{DBA}} & \multicolumn{2}{c|}{\textbf{Neurotoxin}} \\
\cline{2-5}
 & ASR ($\downarrow$) & ACC ($\uparrow$) & ASR ($\downarrow$) & ACC ($\uparrow$) \\
\hline
NoDefense & 95.4\% & 21.2\% & 9.1\% & 92.8\% \\
\hline
FLAME  & 0.0\% & 9.8\% & 0.0\% & 16.8\% \\
FLDetector  & 11.9\% & 85.7\% & 9.6\% & 93.9\% \\
\hline
FL-CLEANER  & 0.2\% & 95.0\% & 0.0\% & 95.1\% \\
\hline
\end{tabular}
\caption{Attack Success Rate (ASR) and Accuracy (ACC) for different defenses and attacks on MNIST. Arrows represent ideal outcome.}
\label{tab:mnistbackdoor}
\end{table}

\begin{table}[h!]
\centering
\setlength{\tabcolsep}{7pt} % adjust column spacing
\renewcommand{\arraystretch}{1.2} % adjust row spacing
\begin{tabular}{|l|c |c|c |c|}
\hline
\multirow{2}{*}{\textbf{Defense}} & \multicolumn{2}{c|}{\textbf{DBA}} & \multicolumn{2}{c|}{\textbf{Neurotoxin}} \\
\cline{2-5}
& ASR ($\downarrow$) & ACC ($\uparrow$) & ASR ($\downarrow$) & ACC ($\uparrow$) \\
\hline
NoDefense & 6.1\% & 65.8\% & 100\% & 59.6\% \\
\hline
FLAME  & 9.1\% & 42.1\% & 23.4\% & 54.1\% \\
FLDetector  & 16.3\% & 75.0\% & 43.9\% & 71.4\% \\
\hline
FL-CLEANER  & 0.2\% & 67.2\% & 0.3\% & 69.1\% \\
\hline
\end{tabular}
\caption{Attack Success Rate (ASR) and Accuracy (ACC) for different defenses and attacks on FashionMNIST. Arrows represent ideal outcome.}
\label{tab:fashionmnistbackdoor}
\end{table}

Concerning backdoor attacks, let us recall that the goal of the attacker is to achieve high ASR while preserving the model's performance on the main task (see Section \ref{subsec:threatmodel}). Thus, we evaluate our defense on the impact it has on this metric. 
To evaluate this, we first test the backdoors in the absence of our defense, and track their ASR over the training epochs. Then, we do the same with a defense applied to compare. In addition, we report the final attacker success rate and test accuracy at the end of the run in separate tables. In our testing, attackers struggled to insert the backdoor when training on the more restrictive custom distribution. So we focused our tests on the Dirichlet distribution.

For comparison with other defenses, we decided not to take into account backdoor defenses the strategy of which includes a tolerance mechanism  sometimes letting one to several attackers get through the defense \cite{fung2020mitigatingsybilsfederatedlearning,rieger_deepsight_2022}. These approaches are by design not byzantine-robust. Indeed, most byzantine attacks only need one attacker to get through. We choose to focus on two backdoor defenses that filter out all detected attackers : FLAME \cite{nguyen2022flame} and FLDetector \cite{zhang2022fldetectordefendingfederatedlearning}. They are also considered as the state-of-the-art \cite{nguyen2023backdoorattacksdefensesfederated}

\begin{figure}[H]
  \centering
\begin{minipage}{.5\textwidth}
\centering
  \includegraphics[width=1\linewidth]{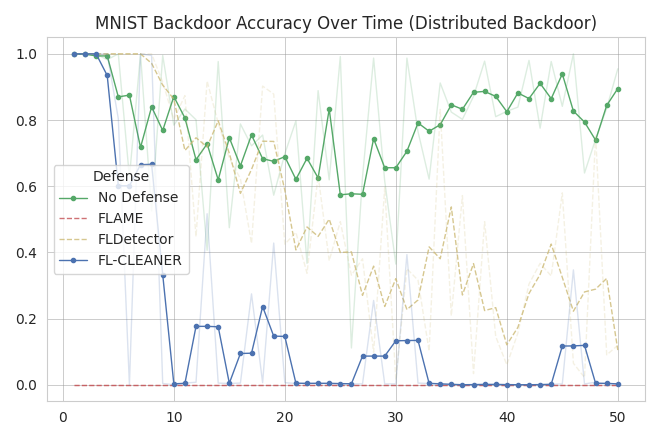}
  \includegraphics[width=1\linewidth]{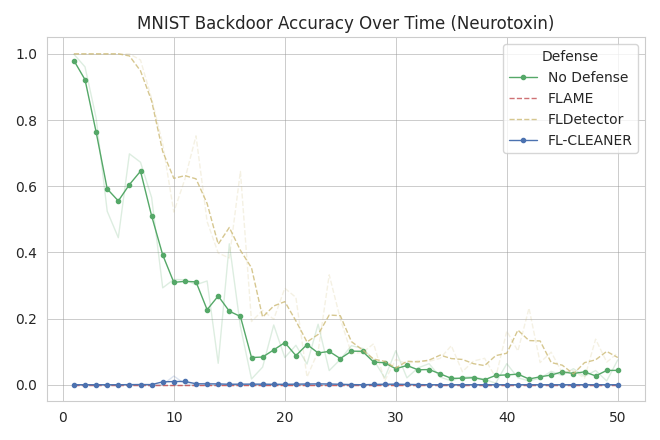}
  \caption{Backdoor Attacker Success rates on MNIST \\ with and without defense (smoothed)}
  \label{fig:fashionmnist_backdoors}
\end{minipage}%
\begin{minipage}{.5\textwidth}
\centering
  \includegraphics[width=1\linewidth]{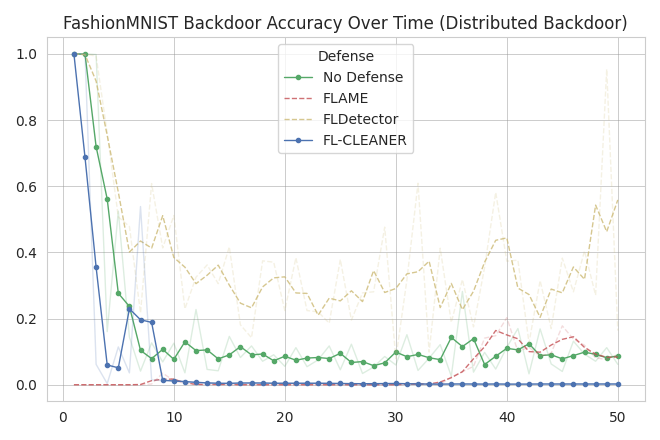}
  \includegraphics[width=1\linewidth]
  {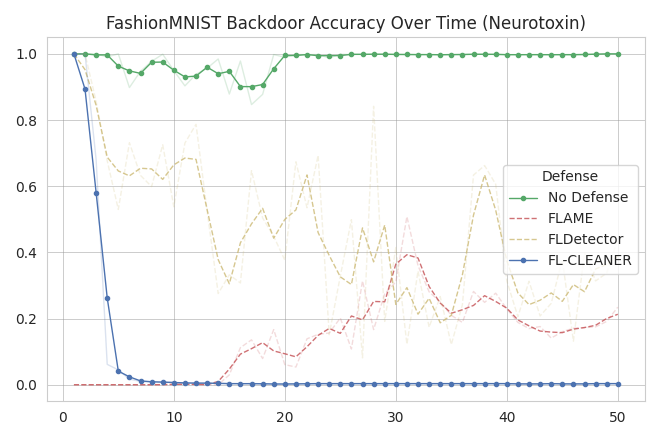}
  \caption{Backdoor Attacker Success rates on Fashion-\\MNIST with and without defense (smoothed)}
  \label{fig:mnist_backdoors}
\end{minipage}%
\end{figure}

Figures \ref{fig:mnist_backdoors} and \ref{fig:fashionmnist_backdoors} show the evolution of backdoor accuracy (ASR) over a 50-round FL run. Tables \ref{tab:mnistbackdoor} and \ref{tab:fashionmnistbackdoor} report the final attacker success rate (ASR) and test accuracy (ACC) observed at the end of these runs. We see that :

\begin{enumerate}[label=\Alph*]
    \item \textbf{FL-CLEANER cancels the effect of both attacks :} After a few rounds, we see that it becomes impossible for the attacker to maintain an ASR not nearing 0\%. This validates the backdoor defense requirement \textbf{R3}.
    \item \textbf{FL-CLEANER is the only effective defense that does not harm the model accuracy :} As seen in Tables \ref{tab:mnistbackdoor} and \ref{tab:fashionmnistbackdoor}, when other defenses manage to mitigate the backdoor attack, it comes at the cost of the model's accuracy. This is due to the defenses blocking too many benign clients. Which makes FL-CLEANER the only defense to validate the defense requirement \textbf{R2}.
\end{enumerate}

\section{Conclusion \& future work}
\label{sec:clc}
In this paper, we introduced FL-CLEANER the first method robust against Byzantine and backdoor attacks under a Non-IID client data distribution. Its originality stands on an autoencoder-based defense system designed to identify malicious clients in Non-IID federated learning environments from updated models' activation maps with other core components: a framework for training CVAEs for an FL defense system leveraging model activation maps normalized with GeoMed, DKL annealing, and a realistic trigger set requirement. Another strength of our proposal is that it depends on a dynamic one-parameter client selection algorithm that minimizes the exclusion of benign clients. We demonstrated that, in image classification benchmarks, FL-CLEANER is robust against Byzantine attacks as well as effective to mitigate tested state-of-the-art backdoor attacks, achieving top performance in benign client misclassification tests compared to current defenses from the literature. Future work will focus on broader machine learning applications, as well as improving our method by enhancing CVAE regularization in the training, and implementing an improved trust propagation algorithm. 
\section*{Acknowledgments}
  This work was partly supported by the Inserm industrial
chair CYBAILE and the PEPR digital health SSF-ML-DH project, ANR-13-
LAB2-0007-01

%Bibliography
\bibliographystyle{unsrt}  
\bibliography{upd}

\end{document}